\documentclass[12pt,preprint]{aastex}

\slugcomment{Accepted for publication in the Astrophysical Journal}

\shorttitle{Evidence for departure from a power-law flare size
distribution}
\shortauthors{Wheatland}

\begin{document}

\title{Evidence for departure from a power-law flare size 
distribution for a small solar active region}

\author{M.~S. Wheatland}
\affil{Sydney Institute for Astronomy, School of Physics, 
  The University of Sydney, NSW 2006, Australia}
\email{m.wheatland@physics.usyd.edu.au}

\begin{abstract}
Active region 11029 was a small, highly flare-productive solar 
active region observed at a time of extremely low solar activity. 
The region produced only small flares:
the largest of the $>70$ Geostationary Observational 
Environmental Satellite (GOES) events for the region has
a peak $1$--$8\mbox{\AA}$ flux of 
$2.2\times 10^{-6}\,{\rm W}\,{\rm m}^{-2}$ (GOES C2.2). The 
background-subtracted GOES peak-flux distribution suggests 
departure from power-law behavior above 
$10^{-6}\,{\rm W}\,{\rm m}^{-2}$, and a Bayesian model comparison
strongly favors a power-law plus rollover model for the distribution 
over a simple power-law model. The departure from the power law 
is attributed to this small active region having a finite amount
of energy. The rate of flaring in the region varies
with time, becoming very high for two days coinciding with the
onset of an increase in complexity of the photospheric magnetic 
field. The observed waiting-time distribution for events is 
consistent with a piecewise-constant 
Poisson model. These results present challenges for models of flare 
statistics and of energy balance in solar active regions.
\end{abstract}


\section{Introduction}

Solar flares are dynamic events in the Sun's outer atmosphere, the
corona, involving the explosive release of magnetic energy in
active regions around sunspots. Large solar flares may produce 
hazardous local space weather conditions but these events are 
relatively infrequent, even at solar 
maximum (e.g.\ Odenwald, Green, \& Taylor 2006; Hudson 2007). 
Small solar flares are much more common, and may also cause 
space-weather effects (e.g.\ Howard \& Tappin 2005; Yermolaev 
et al.\ 2005). Many details of the flare mechanism remain poorly 
understood (e.g.\ Schrijver 2009), although the accepted model is 
magnetic reconnection (e.g.\ Priest \& Forbes 2002). Because 
individual solar flares exhibit great variety, attention has often 
focused on the statistics of multiple events to try to understand the 
phenomenon (e.g.\ Smith \& Smith 1963).

An important statistical property of flares is the appearance of 
a power law in the frequency-size distribution (e.g.\ 
Akabane 1956; Hudson 1969),
i.e.\ the number $f(S)$ of flares observed per unit size $S$ and
per unit time is distributed according to:
\begin{equation}\label{eq:fsize_dist}
f(S)=AS^{-\gamma}.
\end{equation}
By size we mean some measure of the
magnitude of the event, for example the peak radiated flux in a 
short-wavelength band. In equation~(\ref{eq:fsize_dist}) the 
power-law index $\gamma$ is a constant, typically found to be 
in the range $1.5$--$2$ (the exact value depends on the 
chosen measure of size). The factor $A$ determines the mean rate 
of events during an observation period, and may be formally written
\begin{equation}\label{eq:A}
A=\lambda_1 (\gamma -1 )S_1^{\gamma-1},
\end{equation}
where $\lambda_1$ is the mean rate above size $S_1$.

The power-law frequency-energy distribution appears to be universal,
in the sense that the same power-law index is observed from the
Sun at different times (e.g.\ Hudson 1991; Aschwanden, Dennis \& Benz 
1998; Su et al.\ 2006; Christe et al.\ 2008),
although the mean rate of flaring varies substantially. The same
index is also observed for different individual active regions
(Wheatland 2000a). However, an upper limit to the size 
distribution~(\ref{eq:fsize_dist}) is required to ensure that the
mean size of events is finite (assuming $\gamma <2$). Since the 
energy of a flare is expected to scale with size, this represents 
the physical constraint that there is a finite amount of magnetic 
free energy available for flaring. However, it has proven difficult 
to identify this limit in observed frequency-size distributions 
(e.g.\ Hudson 2007), and a departure from the power law has not
been demonstrated for individual active regions. Statistical 
evidence has been presented for an upper limit to the size 
of flares from active regions with small sunspot areas based on
a large sample of small regions (Kucera et al.\ 1997). Proxies for 
the Solar Energetic Particle spectrum also provide indirect 
evidence for a cut-off in the spectrum at large fluences (e.g.\ 
Lingenfelter \& Hudson 1980). The search for departures from the 
flare frequency-size distribution at large event sizes may be 
complicated by detector saturation effects associated with big 
events (e.g.\ Thomson et al.\ 2004).

A second statistical property of flares of interest is the 
waiting-time distribution, or the distribution of times between
flares, which is related to the constant $A$ in 
equations~(\ref{eq:fsize_dist}) and~(\ref{eq:A}). Results for
the waiting-time distribution have proven more ambiguous  
than for the frequency-size distribution, with
a variety of functional forms reported (e.g.\ Pearce et al.\
1993; Biesecker 1994; Wheatland et al.\ 1998; Boffetta et al.\
1999; Moon et al.\ 2001; Wheatland 2001; Moon et al.\ 2002). This
suggests that the distribution is time-dependent (Wheatland \&
Litvinenko 2002) and sensitive to event selection procedures 
(e.g.\ Wheatland 2001; Buchlin et al.\ 2005; Paczuski et 
al.\ 2005; Baiesi et al.\ 2006)

Waiting times in
individual active regions often follow a simple exponential
distribution
corresponding to flares occurring as independent events with
a constant mean rate (a Poisson process in time), or a 
distribution consisting of a sum of exponentials corresponding
to distinct intervals with different rates (e.g.\ Moon et al.\
2001; Wheatland 2001). A suitable model is then a piecewise-constant
Poisson process, with a waiting-time distribution (Wheatland
\& Litvinenko 2002)
\begin{equation}\label{eq:wtd_pconst-poiss}
P(\Delta t)= \sum_i\frac{n_{1i}}{N}\lambda_{1i} 
  {\rm e}^{-\lambda_{1i}\Delta t},
\end{equation}
where $\Delta t$ denotes a waiting time, $n_{1i}=\lambda_{1i} t_i$ 
is the number of events above size $S_1$ corresponding to a rate 
$\lambda_{1i}$ which persists for an interval $t_i$, and where
$N=\sum_i n_{1i}$ is the total number of events. For events over 
longer periods of time including many active regions, a power-law 
tail is observed in the waiting-time distribution (Boffetta et al.\ 
1999). This also may be explained in terms of the time-dependent 
Poisson model (Wheatland 2000b; Wheatland \& Litvinenko 2002), 
although some authors have argued for an intrinsic significance to 
the power law (e.g.\ Lepreti et al.\ 2001).

The observed event distributions have motivated models, in particular 
to account for the power-law frequency-size distribution. A popular
model is the avalanche picture (Lu \& Hamilton 1991; 
Charbonneau et al.\ 2001), which describes a flaring active region
via a cellular automaton model (a field on a grid or lattice) which
is continuously driven and achieves a self-organised critical state.
A flare consists of an avalanche of local energy release 
events which trigger one another. Avalanche models produce a 
power-law frequency-size distribution, where size is the
field energy $E_F$, below an upper rollover defined by the 
finite grid size $L$ (Lu et al.\ 1993; Wheatland \& Sturrock 1996). 
The frequency-energy distribution was
parameterized by Lu et al.\ (1993) using the power-law plus
exponential form:
\begin{equation}\label{eq:avalanche_departure}
f(E_F)\propto E_F^{-\gamma}{\rm e}^{-E_F/E_r},
\end{equation}
where the rollover energy $E_r$ was found to scale as 
$E_r\propto L^{\beta}$, with $\beta \approx 3.9$.
Avalanche models also produce exponential waiting-time 
distributions (e.g.\ Biesecker 1994; Wheatland et al.\ 1998), 
although time-dependent driving alters this (e.g.\ Norman et 
al.\ 2001). 

Another 
approach to modeling flare event statistics involves describing the 
energy balance in a flaring active region in terms of energy input 
and loss (e.g.\ Rosner \& Vaiana 1978; Litvinenko 1994; Craig 2001). 
A general model of this kind was presented 
in Wheatland \& Glukhov (1998), and developed in Wheatland 
(2008; 2009). Active regions are assumed to have a free energy 
$E=E(t)$ which evolves in time $t$ due to deterministic energy input
at a rate $\beta (E,t)$, and due to random downwards jumps 
from energy $E$ to $E^{\prime}$ (flares) at a rate described by 
a transition function $\alpha (E,E^{\prime},t)$. The resulting 
stochastic jump transition model may be formulated either as a 
master equation for the energy distribution $P(E,t)$ (Wheatland 
2008), or as a stochastic differential equation for 
$E(t)$ (Wheatland 2009). In the steady state (i.e.\ for a constant
driving rate and a constant 
total mean rate of flaring), the model can reproduce the observed 
power-law frequency size distribution, below an upper rollover 
defined approximately by the mean energy $\overline{E}$ of the
steady-state distribution $P(E)$. Flares with energy 
$\gg \overline{E}$ are not observed because the active region is 
unlikely to have sufficient energy to produce them. The model 
waiting-time distribution is exponential provided $\overline{E}$ 
is sufficiently large that large flares are unlikely to 
significantly deplete the free energy of the system. In that case 
the total mean rate of flaring is approximately independent of energy, 
and hence does not vary in time, as the active region energy varies.
This produces a Poisson waiting-time distribution. 
The stochastic jump transition model, although idealised, helps to
clarify ideas of energy storage and release in flares, and their
relationship to the flare frequency-energy and waiting-time
distributions.

Studies of flares statistics often use the soft X-ray event lists
generated by the US Space Weather Prediction Center\footnote{See 
http://www.swpc.noaa.gov/.}, which are derived from whole-Sun 
$1$--$8\,\mbox{\AA}$ flux measurements from the Geostationary 
Observational Environmental (GOES) satellites. The peak flux of
GOES events is routinely used to classify flares: very small
flares are labelled A and B class (peak fluxes exceeding 
$10^{-8}\,{\rm W}\,{\rm m}^{-2}$ and $10^{-7}\,{\rm W}\,{\rm m}^{-2}$ 
respectively); small and medium flares are labelled C and M
class (peak fluxes above
$10^{-6}\,{\rm W}\,{\rm m}^{-2}$ and $10^{-5}\,{\rm W}\,{\rm m}^{-2}$ 
respectively); and large flares are labelled X class
(peak flux above $10^{-4}\,{\rm W}\,{\rm m}^{-2}$). A numerical
suffix is used to indicate a multiplicative factor, so that C$3.1$
indicates a peak flux of $3.1\times 10^{-6}\,{\rm W}\,{\rm m}^{-2}$.
The recorded peak fluxes are not background-subtracted.

A recent period of extremely low solar activity (e.g.\ Livingston 
\& Penn 2009; Salabert et al.\ 2009) has afforded a 
unique opportunity to study solar flares occurring in individual
active regions with low levels of background emission from other 
regions on the disk.
In this paper we examine a remarkable sequence of events in active 
region AR 11029 observed from 25 Oct 2009 to 1 Nov 2009.
This small new-cycle sunspot region produced over 70 small soft 
X-ray flares in a dynamic one-week burst of activity, during which 
time it was the only flare-producing 
active region on the Sun. The absence of X-ray emission from other
regions on the Sun at this time
permits careful inspection of the statistics of GOES events for the 
region, including background subtraction of the peak flux.
The key question addressed here is whether a small, highly 
flare-productive active region still exhibits a featureless power 
law in its flare frequency-size distribution, or whether it is 
possible to identify a departure from the power law at large
flare sizes, corresponding perhaps to the region containing a finite
amount of energy. It is also of interest to examine the
waiting-time distribution for such a region, again taking advantage
of the low levels of background emission, to see how the flaring 
rate varies in time.

The sections of the paper are divided as follows. Section~2 
describes the observations of the region. Section~3 presents
analysis, with section 3.1 describing the procedure of background
subtraction of the peak fluxes, and sections 3.2 and 3.3 describing 
the analysis and modeling of the frequency-peak flux distribution
and the waiting-time distribution respectively. Section~4 discussed 
the results, and the appendices present the methods of 
Bayesian inference for a power law and for a power law with an 
upper rollover used in section~3.2.

\section{Observations}

Solar active region AR 11029 emerged on the disk on 21--22 
Oct 2009 and developed sunspots on 24 Oct. Figure~1 shows
a sequence of daily Global Oscillations Network Group (GONG)
magnetograms for that period of time and illustrates the emergence
of the region and its rapid evolution. The region grew in size and 
magnetic complexity as it crossed the disk, arriving at the west limb
on 1 Nov. However the region remained relatively small: the sunspot 
areas recorded in the US National Weather Service/National
Oceanic and Atmospheric Administration (NWS/NOAA) Solar Region 
Summaries\footnote{See http://www.swpc.noaa.gov/.}
are all $<400$ millionths of a hemisphere. Initially the region
had a simple bipolar magnetic configuration and hence was given
the Mount Wilson classification $\beta$. It developed into a 
$\beta$--$\gamma$ region (a more complex bipolar configuration 
such that a single line cannot be drawn to separate the polarities) 
for the period 26 to 29 Oct, and then returned to a 
$\beta$ configuration on 30 Oct.

The photospheric magnetic field of AR 11029 had the polarity 
orientation associated 
with the solar cycle 24, being a northern region with a 
leading negative polarity, although it appeared at an intermediate 
latitude (around 15~deg). At the time of writing (late 2009) there
have been relatively few new cycle regions, and the interval of 
minimum since the peak of the last cycle, around 2001, has been longer 
and quieter than anticipated, with 2008 and 2009 featuring long 
stretches of days without sunspots (e.g.\ Livingston \& Penn 2009;
Salabert et al.\ 2009). 

Although small, AR 11029 was highly flare productive.
The GOES soft X-ray event lists compiled by the US SWPC (discussed
in section~1) list 73 small flares for the interval 24 Oct to 1 Nov.
In the SWPC lists, not all these events are attributed to AR 11029.
However, the active region identification in the GOES event lists 
is quite incomplete: many events lack an associated location and 
active region number. (The identification is dependent on 
ground-based optical observations, and presumably locations are 
missing when there are no available optical observations.)
Inspection of the SolarSoft Latest Events archive produced 
by the Lockheed Martin 
Solar \& Astrophysics Laboratory, which includes Solar and 
Heliospheric Observatory Extreme Ultraviolet Imaging Telescope 
(SOHO/EIT) image sequences for
each event, confirms that all 73 events originate in AR 
11029.\footnote{The Latest Events data is available at
http://www.lmsal.com/solarsoft/latest\_events/.} The flares are all
small (one A-class event, 60 B-class events, and 11 C-class events), 
with the largest being C2.2. Table~1 summarises the daily
behavior of the active region, and lists the numbers of flares 
observed per day. The region was particularly flare productive 
on 26 Oct and 27 Oct, the first two days of assignment of the 
magnetic classification $\beta$--$\gamma$.

Figure~2 illustrates the activity observed in AR 11029. The upper 
panel shows the one-minute GOES $1$--$8\,\mbox{\AA}$ 
flux values versus time from 23 Oct to 3 Nov, in a log-linear
representation. Prior to the emergence of 
AR 11029 the X-ray flux was at the base level for the detector, 
and it returned to this value when the region rotated off the 
disk. This suggests that all $1$--$8\,\mbox{\AA}$ emission 
from the Sun in this interval originates from AR 11029, providing 
observation conditions which are close to ideal for
identification of X-ray events from this region. The top panel of 
Figure~2 may be regarded as a 
soft X-ray light-curve for AR 11029. The lower panel of Figure~2
illustrates the GOES events identified by the US SWPC for the
region, in a schematic
diagram showing the peak flux of each event versus the peak 
time, with a vertical line drawn for each event. The
event lists provide a start time, a peak time, and an end time 
for each event, as well as a peak flux. 
In the analysis in this paper the peak time 
is used to represent the event time. The lower panel of Figure~2
shows an initial interval with a low flaring rate (prior to
26 Oct), two days of intense flaring (26 Oct to 27 Oct,
inclusive), and then a third interval with a low flaring 
rate.

\section{Analysis}

\subsection{Background subtraction}

The upper panel of Figure~2 illustrates issues associated with
GOES event selection and background emission (Wheatland 2001).
Even for a small active region observed at a time when there are
no other X-ray emitting regions on the disk, a period of flaring
is seen to raise the soft X-ray background by more than an order 
of magnitude. The SWPC events in the lower panel are selected
against the time-varying background, and the peak fluxes shown 
are not background subtracted. The time-varying background leads 
to incompleteness in the event lists. The start of a 
GOES event is defined\footnote{See descriptive text at 
http://www.ngdc.noaa.gov/stp/SOLAR/ftpsolarflares.html.} by 
four monotonically increasing one-minute flux values 
that produce a $40\%$ enhancement over the initial flux. At times 
of high background flux, small events fail to produce a $40\%$ 
enhancement over the background and are omitted.
The absence of background-subtraction leads to time-dependent
event definitions, so that events with the same intrinsic peak flux
are assigned different peak fluxes at different times.

The absence of background subtraction for the peak fluxes is 
particularly important for small events, in which case the 
background is a substantial contributor to the event peak
flux. To study the frequency-peak flux distribution, we 
background subtract each peak flux, using the
one-minute GOES flux values.\footnote{The Interactive Data Language 
(IDL) SolarSoft GOES routines -- see http://www.lmsal.com/solarsoft/
--- are used to obtain the data.} For each event the flux values
prior to the start time $t_{\rm s}$ are averaged over an interval
equal to the rise time $t_{\rm p}-t_{\rm s}$ of the event, where
$t_{\rm p}$ is the peak time. The resulting flux is taken as the 
background 
estimate, and is subtracted off the peak flux value for the analysis
in sections~3.2 and~3.3. This method of background subtraction 
is a suitable simple method. 
Any approach to background subtraction faces the problem of (unknown) 
variation in the background between the start time of an event and 
the peak time. The chosen approach is intended
to capture typical variation over the relevant time scale for 
which the background is unknown (the rise time), and to provide a 
representative background flux.

Figure~3 illustrates the procedure and the results. The upper
panel shows the first event from AR 11029, which occurred around 
20:00~UT on 24 Oct. The one-minute flux values are shown versus 
time (solid curve), with the three vertical 
dotted lines indicating $t_{\rm s}$, 
$t_{\rm p}$, and $t_{\rm e}$ (the end time) for the event, as given
by the SWPC event lists. The averaging interval is the
interval to the left of the first vertical line, and 
the resulting background estimate is shown by the dotted horizontal 
line. This event is assigned a peak flux 
$1.6\times 10^{-7}\,{\rm W}\,{\rm m}^{-2}$ in the SWPC event lists
(B1.6). The background estimate is 
$3.11\times 10^{-8}\,{\rm W}\,{\rm m}^{-2}$, so the resulting
background-subtracted peak flux is 
$1.29\times 10^{-7}\,{\rm W}\,{\rm m}^{-2}$ (B1.3). The lower 
panel of Figure~3 shows the effect of 
background subtraction for all events, displayed as a cumulative 
number distribution, i.e.\ the number of events with a 
larger (or equal) peak flux, versus peak flux. The plot
is shown in a log-log representation. The peak-flux values before
background subtraction are indicated by the crosses, and the 
background-subtracted values are shown by the diamonds. The
procedure introduces substantial changes in the peak-flux
values, in particular for smaller events. 

\subsection{The frequency-peak flux distribution}

The cumulative number distribution shown in the lower panel of
Figure~3 is expected to follow a power law, according to 
equation~(\ref{eq:fsize_dist}). Specifically, the model cumulative 
number distribution corresponding to the frequency-size distribution
equation~(\ref{eq:fsize_dist}) is
\begin{equation}
C(S)=N_1\left(S/S_1\right)^{-\gamma+1},
\end{equation}
where $N_1=\lambda_1 T$ is the number of events above
size $S_1$ in time $T$. Inspection of Figure~3 suggests an
absence of large events by comparion with the power-law
form, for the background-subtracted distribution -- the distribution
falls away rapidly at large $S$. To quantify this behavior, we
consider comparison of the observed (background-subtracted) 
distribution with two models:
a simple power-law, and a power law with an upper exponential rollover.
The probability distributions for peak flux for the two models 
are
\begin{equation}\label{eq:PS_pl}
P_{\rm pl}(S) = (\gamma_{\rm pl}-1)S_1^{\gamma_{\rm pl}-1}S^{-\gamma_{\rm pl}},
\end{equation} 
and
\begin{equation}\label{eq:PS_plr}
P_{\rm plr}(S) = BS^{-\gamma_{\rm plr}}{\rm e}^{-S/\sigma }, 
\end{equation}
respectively, with $S\geq S_1$. In equation~(\ref{eq:PS_plr}) 
$\sigma $ is the upper rollover and $B$ is a
normalization constant, which depends on $\gamma_{\rm plr}$, 
$\sigma$, and $S_1$ (see Appendix B). The frequency-peak flux 
distribution is obtained from the probability distribution for
peak flux by multiplying by the rate $\lambda_1$.

Equation~(\ref{eq:PS_plr}) 
provides a simple model with an upper departure from a power law and 
is not a physically-motivated model as such, although as discussed in
section~1 this functional form was used to represent finite grid-size 
effects in the avalanche models (Lu et al.\ 1993; Wheatland \& Sturrock 
1996). The model is chosen here for its simplicity. Other models
are possible, and are not intended to be excluded by this choice: 
for example a broken power-law model could be used. The important 
aspect of the model for the analysis presented here is that it includes 
a departure from power-law behavior at large event sizes.

The appendices give the details of the Bayesian methods used 
to estimate parameters and to compare the two model 
distributions~(\ref{eq:PS_pl}) and~(\ref{eq:PS_plr}) against
the data. Bayesian probability provides a general
framework for inference in science (see e.g.\ Jeffreys 1961; 
Jaynes 2003; Sivia \& Skilling 2006). The advantages of the Bayesian
methods used here are that they avoid binning the data, and so exploit 
all information in a small data set. The parameter estimation 
methods generalise a maximum likelihood method for estimation of 
a power-law index which is commonly used in astrophysics 
(Crawford, Jauncey, \& Murdoch 1970; Bai 1993; Wheatland 2004). 
The model comparison presented in Appendix~C is new, but follows 
the standard Bayesian approach.

Figure~4 illustrates the 
data and the analysis. The 
upper panel shows the observed number distribution for the 
background-subtracted peak flux, plotted as a histogram 
with logarithmic binning in peak flux, and the lower panel
shows the cumulative number distribution following Figure~3. 
Both panels use log-log representations. The ordinate values for 
the histogram (upper panel) 
are the numbers of events in each bin divided by the bin 
width, so the histogram corresponds to $N(S)=N_1P(S)$, where $N_1$ 
is the total number of events above size $S_1$ and $P(S)$ is the 
probability 
distribution for peak flux. The vertical dotted line in each panel
shows the value of the lower peak-flux limit 
$S_1=10^{-7}\,{\rm W}\,{\rm m}^{-2}$ assumed for the models.
The value $S_1$ is chosen by inspection
of the upper panel of Figure~2. This choice represents a value for
flux which is larger than the background over most of the period 
of observation, excluding during brief intervals. As such,  
all events of this size or larger are expected to be identified and
included in the SWPC event lists. This choice gives $N_1=56$.
The histogram in Figure~4 (upper panel) reveals approximate
power-law behavior over a limited range, as expected. The
error bars shown correspond to the square root 
of the number of events in each bin, and are intended to be 
indicative values only (they are not used in the analysis). 
The binned event numbers are also not used in the analysis and
the upper panel of Figure~4 is intended only for illustration.
The Bayesian methods used here explicitly include every data 
point in the
analysis, which is important given the small number of events.

Figure~4 shows the model comparison using the Bayesian
methods described in the appendices. The solid
lines in the two panels of Figure~4 represent the
simple power-law model, and the solid curves represent the 
power-law plus rollover model. The power law
index for the simple model is $\gamma_{\rm pl}=1.88\pm 0.12$, 
where the estimate and the uncertainty are obtained from moments
of the posterior distribution for 
$\gamma$ (see Appendix A). The power-law index for the
power-law plus rollover model is $\gamma_{\rm plr}=0.99\pm 0.34$, 
and the rollover value is 
$\sigma = 9.8\pm 5.3\times 10^{-7}\,{\rm W}\,{\rm m}^{-2}$. The
estimates and uncertainties correspond to moments of the
marginal posterior distributions for $\gamma$ and $\sigma$
(Appendix B). The uncertainties are quite large, reflecting the
difficulty of inferring two parameters from a small data set.
However, the lower panel of Figure~4 indicates that the power-law 
plus rollover model provides a much better fit to the data than 
the simple power-law model.

To quantify the comparison of the models, we consider the 
ratio of their posterior probabilities, a 
standard Bayesian procedure (e.g.\ Jaynes 
2003; Gregory 2005; Sivia \& Skilling 2006). According to 
Bayes's theorem the ratio of the posterior probabilities for
the models, which is called the odds ratio, is
\begin{equation}\label{eq:odds_ratio_body}
r_{\rm plr/pl}(D)
=\frac{P(D|{\cal M}_{\rm plr})}{P(D|{\cal M}_{\rm pl})}
\times \rho_{\rm plr/pl},
\end{equation}
where the simple power law and power-law plus rollover models 
are denoted ${\cal M}_{\rm pl}$ and ${\cal M}_{\rm plr}$,
respectively, where $D$ denotes the data, and where $\rho_{\rm plr/pl}$
is the prior odds ratio 
\begin{equation}\label{eq:prior_odds_body}
\rho_{\rm plr/pl}=\frac{P({\cal M}_{\rm plr})}{P({\cal M}_{\rm pl})}.
\end{equation}
The the prior odds ratio represents
what was thought about the relative probabilities of the two
models before looking at the data. The other term on the right hand 
side of equation~(\ref{eq:odds_ratio_body}) is the `Bayes factor'
$P(D|{\cal M}_{\rm plr})/P(D|{\cal M}_{\rm pl })$, which is
the ratio of the likelihoods of the models. This 
represents what the data says about the models.  The two models 
have different numbers of free parameters, and the Bayesian 
approach takes this into account by expanding the likelihoods in
the Bayes factor as 
integrals over the possible parameter values, so that, for example, 
the likelihood for the simple power law model is expressed as
\begin{equation}\label{eq:global_like_pl_body}
P(D|{\cal M}_{\rm pl })= \int d\gamma P(D|\gamma ) P(\gamma ),
\end{equation}
where $P(D|\gamma )$ is the likelihood for a given $\gamma$ and
$P(\gamma )$ is the prior distribution for $\gamma$ (see 
appendices for full details).
Model likelihoods expressed in this way are often called global 
likelihoods. Appendix~C provides expressions for the global
likelihoods $P(D|{\cal M}_{\rm pl})$ and
$P(D|{\cal M}_{\rm pl})$ for the two models, which permit the 
evaluation of the Bayes factor and hence the odds 
ratio via~(\ref{eq:odds_ratio_body}).

For the 56 GOES peak fluxes associated with active region 11029
and above size $S_1$
the odds ratio evaluates to $r_{\rm plr/pl}(D)\approx 220$, assuming
a unity prior odds ratio ($\rho_{\rm plr/pl}=1$).
Odds ratios are often presented in decibels (dB) and in this 
case the odds ratio in decibels is 
$10\log_{10} r_{\rm plr/pl}(D)\approx 23\,{\rm dB}$. 
This result implies that, if both models are assumed a priori
to be equally likely, then the data favors the power-law plus 
rollover model by a factor of more than 200. An odds ratio of 
this magnitude may be interpreted as strong evidence for the 
favored model (e.g.\ Jeffreys 1961; Jaynes 2003; see also the 
discussion at the end of Appendix~C), and the result confirms 
the qualitative impression 
given by the lower panel of Figure~4. Of course, if the simple 
power-law model is thought a priori to be strongly preferable 
($\rho_{\rm plr/pl} \ll 1$), then the odds ratio is reduced,
according to equation~(\ref{eq:odds_ratio_body}). 
However, at face value the data for active region 11029 
provides strong evidence for the power-law plus rollover model 
over the simple power-law model. 

The specific value of the odds ratio depends on the 
choice of the threshold $S_1$. However, the result does not 
depend strongly on this choice in the sense that the model 
with the rollover is always favored. If this choice is made 
too large the odds ratio is reduced due to the resulting small 
event numbers.

The power-law plus rollover model is chosen for simplicity,
and it is worthwhile to consider other possible models, for 
example a broken power law. It is unlikely that the small set of 
data for AR 11029 allows distinction between a broken 
power-law model and a power law with an exponential rollover, and 
the calculation is not attempted. However, because the Bayesian 
model comparison involves integration over all possible values of 
the rollover parameter $\sigma$, the specific form of the model is 
likely to be less important in the model comparison than the 
choice of a model with a departure from power-law behavior. 
It is expected that a 
model comparison between a simple power law and a broken power 
law will give very comparable results for this data (i.e.\ 
an odds ratio strongly in favor of the broken power law).

\subsection{The waiting-time distribution}
 
The waiting-time distribution is also constructed for the 56
events with background-subtracted peak flux larger than 
$S_1=10^{-7}\,{\rm W}\,{\rm m}^{-2}$, to investigate the flaring 
rate. Figure~5 illustrates the analysis. 
The upper panel shows the cumulative number of events versus time,
the middle panel shows the Bayesian blocks analysis of the rate
versus time (Scargle 1998), and the lower panel shows the 
waiting-time distribution
(diamonds), together with the piecewise constant Poisson model
produced by the Bayesian blocks procedure (solid curve). 

The upper panel of Figure~5 indicates the rate, qualitatively, by
the slope of the cumulative number versus time plot. The rate is
clearly low (up to 26 Oct), then high (26 Oct and 27 Oct), and then
low again. This matches the variation seen in the daily event 
numbers in Table~1 for all events before background subtraction.

The middle panel of Figure~5 shows the Bayesian blocks procedure 
due to Scargle
(1998) applied to the background-subtracted data. This procedure
takes a time history of events and determines a most probable
piecewise constant Poisson model using iterated Bayesian hypothesis 
testing. 
The resulting intervals with (approximately) constant rates are 
called Bayesian blocks. As shown in the panel, three blocks are
identified by the procedure, matching the qualitative impression
given by the upper panel. The three rates (for events above size
$S_1$) are $\lambda_{11}=0.17\,{\rm hr}^{-1}$, 
$\lambda_{12}=0.97\,{\rm hr}^{-1}$, and
$\lambda_{13}=0.086\,{\rm hr}^{-1}$.

The lower panel of Figure~5 shows the observed waiting-time 
distribution (diamonds),
in a log-linear representation, together with the 
model defined by equation~(\ref{eq:wtd_pconst-poiss}) with the three
rates and intervals identified by the Bayesian blocks procedure 
(solid curve). The distribution appears as a double exponential.
This may be explained qualitatively in terms of a Poisson process
with two different rates [see equation~(\ref{eq:wtd_pconst-poiss})],
which is appropriate because the two intervals with low rate have 
similar values for the rate. The quantitative result obtained with
the piecewise constant model (solid curve) is seen to reproduce 
the observed waiting-time distribution.

\section{Discussion}

Active region AR 11029 is an example of a small, highly 
flare-productive active region. Its appearance at a time
of extremely low solar activity provides an opportunity to examine
soft X-ray flare production for an individual region against a 
very low background, and hence to look carefully at the 
statistics of flare occurrence. The observing conditions are close
to ideal for identification of soft X-ray events from one region.
The event lists produced by the 
US Space Weather Prediction Center (SWPC) from Geostationary 
Observational Environmental Satellite (GOES) data are used. A
total of 73 soft X-ray flares are listed for the interval the 
region was on the disk (21 Oct to 1 Nov 2009). All of the events 
are small, with the largest having a peak flux of 
$2.2\times 10^{-6}\,{\rm W}\,{\rm m}^{-2}$ (C2.2).
The Lockheed Martin Latest Events catalog demonstrates that 
all events are produced by AR 11029, a remarkable instance of 
flaring from one region in the absence of other activity. The 
events are here subject to individual background subtraction, and 
the flare frequency-peak flux, and waiting-time distributions are 
analyzed.

The frequency-peak flux distribution for the background-subtracted 
events reveals an absence of large
events by comparison with the expected power-law form. The deficiency
is clearly seen in a cumulative number representation of the 
distribution (lower panels of Figures~3 and~4). A quantitative comparison
is made with a simple power-law model, and a power-law plus upper
exponential rollover model, using Bayesian methods, outlined in
the appendices. The power-law plus rollover model is strongly favored
by the data: the odds ratio for the two models is $\approx 220$, or 
$23\,{\rm dB}$, for a prior odds ratio of unity. This means that, if
the two models are assumed a priori to be equally likely, then the
power-law plus rollover model is more than $200$ times more probable,
based on the data. The model comparison takes into account all 
possible values of the model parameters. The result represents strong 
evidence in favor of a departure from power-law behavior in the 
frequency-peak flux distribution. As discussed in section~3.2, the
power-law plus rollover model is chosen for convenience, and other
models (e.g.\ a broken power law) are also possible. It is expected
that a broken power law model would also be strongly favored over
a simple power law for this data. The important result is that the
events shows departure from the power law peak-flux distribution at
large peak flux.

The waiting-time distribution for the background-subtracted events
(lower panel of Figure~5)
has a double-exponential form, which may be explained via a 
piecewise-constant
Poisson model with two different rates. The active region initially
produced flares at a low rate (prior to 26 Oct), then at a high
rate (26 Oct and 27 Oct), and then at a low rate again, with the 
second low rate comparable to the first low rate. A Bayesian analysis
identifies the three intervals, and the model piecewise-constant 
Poisson waiting-time distribution produced by the analysis reproduces
the observed waiting-time distribution.

The reported departure of the frequency-peak flux distribution from 
the simple power-law form is the first time such a 
result has been seen for an individual active region. 
Evidence for a size-limit was reported for a large statistical
sample of small solar active regions (Kucera et al.\ 1997), and a 
departure is expected on energetics grounds (e.g.\ Hudson 1991), 
but in general it has proven difficult to identify the effect in data 
(e.g.\ Wheatland 2000a; Hudson 2007). This study has certain
advantages over previous studies. First, a small region is observed 
at a time of very low solar activity, leading to GOES event lists 
for the region which are particularly complete above the chosen 
peak-flux threshold. The low background also allows careful 
background subtraction of each event to 
determine the intrinsic peak flux. The lower panel of Figure~3 
suggests the importance of background subtraction for the 
analysis. Previous studies of flare statistics, in particular 
using the GOES events, have suffered from bias due to the problem 
of event selection against a large time-varying background (e.g.\ 
Wheatland 2001). A second advantage of this study is the 
very high flare-productivity of this small region. The large
number of events means that the peak-flux distribution is 
well-defined, including for large peak fluxes.
A third advantage is the application of Bayesian statistical 
techniques which count every event equally. The methods outlined 
in the appendices avoid binning the data and hence use all 
available information.

The simplest interpretation of the observed departure from a power 
law in the peak-flux distribution is that it reflects the finite
magnetic free energy available for flaring in a small region
(the peak flux may be considered a proxy for energy). As discussed
in section~1, a departure from power-law behavior is required on
energetics grounds for any active region and appears in models 
for flare statistics. For the avalanche models 
(e.g.\ Lu \& Hamilton 1991; Charbonneau et al.\ 2001) it
corresponds to the finite cellular automata grid, and for
the energy balance models (Wheatland 2008; 2009) it is defined by
the largest energy the system is likely to attain. However, the 
models are typically assumed to operate in a regime where the 
free energy of the system exceeds the typical energy of flares,
so that a power-law size distribution is generated over many
decades in size. In other words, they model large active regions,
for which the finite-energy effect is not observed. The results 
reported for AR 11029 may lead to new ways to test the statistical 
models.

It is interesting to relate the peak-flux value of the departure
from power-law behavior observed for this region to the size of 
the region, and to the largest flares observed in big active
regions. One of the larger 
flares of solar cycle 23 was the X4.8 event in AR 10039 on 23 Jul
2003 (e.g.\ Emslie et al.\ 2004). According to the Solar Region 
Summaries prepared by the US National
Weather Service/National Oceanic and Atmospheric Administration
(NWS/NOAA) this region had a size of at least 
$1000\,\mu$-hemispheres at the time of the event, and
so was perhaps five times larger than AR 11029. However, this
is the spot area alone, and the area of enhanced magnetic field
is likely to be much greater, so we consider the ratio of the 
relevant areas of the two regions to be a factor of $10$.
The avalanche
model scaling of rollover energy $E_r$ and grid size found by
Lu et al.\ (1993)
[see equation~(\ref{eq:avalanche_departure})] implies a
dependence on area of $E_r\propto A^{\beta/2}$, with 
$\beta\approx 3.9$. Assuming the 23 Jul 2003 event corresponds
roughly to a maximum-energy event for AR 10039, and assuming that
event energy is proportional to peak X-ray flux (e.g.\ Lee et al.\ 
1995), leads to a rough prediction for a maximum peak flux for the
small active region AR 11029 of 
$\approx 4.8\times 10^{-4}\,{\rm W}\,{\rm m}^{-2}
  \times 10^{-\beta/2}\approx 5\times 10^{-6}\,{\rm W}\,{\rm m}^{-2}$,
i.e.\ a C5 event, which is roughly consistent with the 
observations.

The results for the waiting-time distribution confirm simple
Poisson occurrence of flares in time, and the tendency of active
regions to exhibit intervals with enhanced flaring rates. The
interval with a higher rate corresponds to the start of the 
interval when the active region is assigned a more complex 
photospheric magnetic field configuration (Mt Wilson classification
$\beta$--$\gamma$).
It is well known that more complex magnetic configurations are
associated with higher flaring rates (e.g.\ Sammis et al.\ 2000),
although the underlying physical mechanisms are poorly understood. 
However, the flaring rate also returns to a lower rate during the 
interval for which the active region had the $\beta$--$\gamma$
classification, suggesting also the difficulty 
of flare prediction based on photospheric magnetic field data 
alone (e.g.\ Barnes \& Leka 2008).

The avalanche models, and the energy balance model due to 
Wheatland \& Glukhov (1998) and Wheatland (2008; 2009) predict 
a simple exponential waiting-time distribution in the steady state 
(i.e.\ when the driving rates and mean flaring rates are 
constant). In the avalanche 
model this corresponds to the system being equally likely to 
avalanche at any time due to the self-organised critical
state. This state is dependent on the total energy being much 
larger than the energy of events. If events significantly deplete 
the free energy of the system, departure from an exponential 
waiting-time distribution is expected, as the system requires
some time to re-establish the self-organised critical state 
when a very large flare 
occurs. Similarly, for the energy balance models in a 
steady-state, the total flaring rate is independent of the energy 
of the system, and hence is time-independent, provided the 
mean energy of the system is much greater than the energy of
the largest flares 
(Wheatland 2008; 2009). A Poisson waiting-time distribution 
is obtained if individual events do not significantly deplete 
the energy, 
but departures from Poisson behavior are observed if
events do reduce the overall energy. Hence the results 
presented here also challenge the statistical models: the 
observed frequency-size distribution 
departs from a power law, implying that flares are depleting the
energy substantially, but the Poisson model accounts for the
observed waiting-time distribution (Figure~5). The models should
be re-considered in this light. This brief discussion 
neglects the influence of a time-dependent flaring rate 
on the waiting-time distribution, and this needs also 
to be considered. 

This paper demonstrates the utility of flare 
statistics for providing insight into the flare mechanism, and 
also the need for consideration of potential biases in studies.
Many studies of flare statistics suffer from bias associated
with event definition and selection, and the problems tend to be 
more severe at times of high solar activity, when there are more
flare-producing active regions on the Sun. The GOES event lists
present particular difficulties due to the time-varying
soft X-ray background generated by multiple flare occurrence 
(Wheatland 2001), but the lists have been widely used to 
investigate flare statistics because of their automatic 
availability and the extensive archive of observations, which 
spans three solar cycles (the event lists data back to 
1975).\footnote{Archives accessible at 
http://www.ngdc.noaa.gov/stp/SOLAR/ftpsolarflares.html.} It is 
important to be aware of these limitations when using GOES data to 
study flare statistics. This study has focused on a single region
observed at a time of extremely low solar activity, which reduces 
the problem, and the individual peak fluxes have been 
background-subtracted.

The evidence for departure from a power-law flare size distribution 
presented here is strong,
but it is based on a single region and data set. The results
should be confirmed by similar investigation of other small 
regions, and comparison with data from other instruments in 
follow-up studies. 

\acknowledgments

Mike Wheatland acknowledges the work of Tomonori Hu on one of 
the codes, and Don Melrose for comments on drafts. Thoughtful 
comments by an anonymous referee also have helped to improve 
the paper, and we thank the referee in particular for 
the estimate relating maximum peak flux and region area 
given in section~4.

{\it Facilities:} \facility{GOES}, \facility{SOHO},
\facility{GONG}.

\section*{Appendices}

\appendix

In the following we outline Bayesian methods of inference
for a power-law model, and for a power law with an
upper exponential rollover. The methods in sections B and C 
are new, and should be of general use in astrophysics.

\section{Inference on a power law}

Crawford, Jauncey, \& Murdoch (1970) and Bai (1993) 
describe a maximium likelihood approach to estimating a 
power-law index from data. A Bayesian version was presented
in Wheatland (2004), including a simple uncertainty estimate. 
That approach is extended here to include model comparison.

The model probability distribution is
\begin{equation}\label{eq:PS_pl_appA}
P (S) = (\gamma-1)S_1^{\gamma -1}S^{-\gamma },
\end{equation} 
with $S\geq S_1$. The likelihood function, that is 
the probability of observed data $D=\{s_1,s_2,...,s_M\}$
given the model, is the product of the probabilities of each 
datum implied by equation~(\ref{eq:PS_pl_appA}):
\begin{equation}\label{eq:pl_likelihood}
P(D | \gamma )
  = \left[(\gamma-1)/S_1 \right]^{M}\pi^{-\gamma}(ds)^M,
\end{equation}
where
\begin{equation}
\pi\equiv \prod_{i=1}^{M}s_i/S_1,
\end{equation}
and where $s_i\geq S_1$ for each $i$.
Bayes's theorem provides the 
posterior distribution for $\gamma$, that is the
probability of the model given the data: 
\begin{equation}\label{eq:bayes}
P(\gamma | D )=
   P(D | \gamma ) P(\gamma )/P(D)
\end{equation}
where $P(\gamma  )$ is the prior distribution for
$\gamma$ (the distribution assigned to $\gamma$ in
the absence of the data), and $P(D)$ is a term which does not
depend on the model parameter $\gamma$. For the purposes of parameter
estimation $P(D)$ is determined by 
normalisation, i.e.\ by requiring that the integral of the posterior
over all possible values of $\gamma$ is unity. In the following 
we use a uniform prior: 
\begin{equation}\label{eq:pl_prior}
P(\gamma ) = \left\{
  \begin{array}{ll}
  (\gamma_2-\gamma_1)^{-1} & 
  \mbox{if $\gamma_1\leq \gamma \leq \gamma_2$}
  \\
  0 & \mbox{else.}
\end{array}
\right.
\end{equation}
Equations~(\ref{eq:pl_likelihood})--(\ref{eq:pl_prior}) give
the posterior distribution for
$\gamma$:
\begin{equation}\label{eq:prob_gam}
P(\gamma|D )= C \frac{(\gamma-1)^{M}}{\pi^{\gamma}}
  P (\gamma),
\end{equation}
where $C$ is the normalization constant, which for a uniform prior is
\begin{equation}\label{eq:norm_pl}
C=\frac{(\gamma_2-\gamma_1) \pi (\ln \pi )^{M+1}/M!}
  {P[M+1,(\gamma_2-1)\ln\pi ]
  - P[M+1, (\gamma_1-1)\ln \pi ] }.
\end{equation}
In equation~(\ref{eq:norm_pl}) $P (a,x)$ denotes the incomplete 
Gamma function:
\begin{equation}\label{eq:P(a,x)}
P(a,x)=\frac{1}{\Gamma(a)}\int_0^x {\rm e}^{-t}t^{a-1}dt,
\end{equation}
and
\begin{equation}\label{eq:Gamma)}
\Gamma (a)=\int_0^{\infty}{\rm e}^{-t}t^{a-1}dt
\end{equation}
is the Gamma (factorial) function  (Abramowitz \& Stegun 1964).

The posterior distribution contains all the information available
for inference, and different best estimates for the power-law index
may be extracted from this distribution. 
The most probable value of $\gamma$ is the maximum 
of $P(\gamma|D)$, which is called the modal estimate: 
\begin{equation}\label{eq:gam_ML}
\gamma_{\rm mod}=\frac{M}{\ln\pi}+1.
\end{equation}
This corresponds to the maximum likelihood estimate of $\gamma$ 
given by Crawford, Jauncey, \& Murdoch (1970), and Bai (1993). 
A simple corresponding
estimate of the uncertainty in the 
most likely value of $\gamma$ (for a uniform prior) was given by 
Wheatland 
(2004) using the assumption of Gaussian behavior of $P(\gamma|D )$ 
in the vicinity of the peak (e.g.\ Sivia \& Skilling 2006). In that 
case the width of the distribution is roughly 
$\sigma_{\rm mod}=[L^{\prime\prime}(\gamma_{\rm mod})]^{-1/2}$, 
where $L(\gamma)=-\ln \left[ P(\gamma|D )\right]$ and where the 
prime denotes a derivative. This leads to 
$\sigma_{\rm mod}=M^{1/2}/\ln\pi$. Using equation~(\ref{eq:gam_ML}) 
then gives 
\begin{equation}\label{eq:sig_gam_Gauss}
\sigma_{\rm mod}= \frac{\gamma_{\rm mod}-1}{M^{1/2}}.
\end{equation}

An alternative (and more general) approach to estimation is 
provided by moments of the posterior distribution. Specifically,
if the $k^{\rm th}$ moment is defined by
\begin{equation}
\gamma^{[k]}=\int d\gamma \,\gamma^k P(\gamma |D)
\end{equation}
then a best estimate and a corresponding uncertainty are provided
by
\begin{equation}\label{eq:est_mom}
\gamma_{\rm mom}=\gamma^{[1]}
\end{equation}
and
\begin{equation}
\sigma_{\rm mom}^2=\gamma^{[2]}-\left(\gamma^{[1]}\right)^2.
\end{equation}

\section{Inference on a power law with a rollover}

The model probability distribution is 
\begin{equation}\label{eq:PS_plr_rep}
P(S) = BS^{-\gamma}{\rm e}^{-S/\sigma }, 
\end{equation}
with $S\geq S_1$, where $\sigma $ is the upper rollover, and where
the normalisation constant is
\begin{equation}
B=\frac{\sigma^{\gamma-1}}{\Gamma (1-\gamma)
\left[1-P(1-\gamma,S_1/\sigma )\right]}.
\end{equation}
In this case the likelihood function for the data 
$D=\{s_1,s_2,...,s_M\}$ is
\begin{equation}\label{eq:lik_uroll}
P(D|\sigma,\gamma ) =
  \frac{\sigma^{M(\gamma-1)}S_1^{-M\gamma}
  \pi^{-\gamma}{\rm e}^{-S_1\Sigma/\sigma}(ds)^M}
  {\left[\Gamma(1-\gamma)
  \left\{1-P(1-\gamma,S_1/\sigma )\right\}\right]^M},
\end{equation}
where
\begin{equation}
\Sigma=\sum_{i=1}^Ms_i/S_1.
\end{equation}
Bayes's theorem may be stated as
\begin{equation}
P(\sigma, \gamma |D )
=P(D|\sigma ,\gamma )P(\sigma )P(\gamma )/P(D),
\end{equation}
where $P(\sigma, \gamma |D )$ is the joint posterior distribution 
for $\sigma$ and $\gamma$, and $P(\sigma )$ and $P(\gamma )$ are
the prior distributions for the two parameters. 
Following the approach for the simple power law, we choose a uniform
prior for $\gamma$, given by equation~(\ref{eq:pl_prior}). However, 
the rollover $\sigma$ appears in equation~(\ref{eq:PS_plr_rep}) as 
a scale factor, in which case it is appropriate to use a 
Jeffreys prior (Jeffreys 1961; Jaynes 2003):
\begin{equation}\label{eq:sigma_prior}
P(\sigma )=\frac{1}{\sigma }.
\end{equation}
With this choice and with the uniform prior for $\gamma$ the joint
posterior distribution evaluates to 
\begin{equation}\label{eq:Ppost_uroll}
P(\sigma,\gamma|D) =
  \frac{E\sigma^{M(\gamma-1)-1}S_1^{-M\gamma}
  \pi^{-\gamma}{\rm e}^{-S_1\Sigma/\sigma}}
  {(\gamma_2-\gamma_1)
  \left[\Gamma(1-\gamma)
  \left\{1-P(1-\gamma,S_1/\sigma )\right\}\right]^M},
\end{equation}
where $E$ is a normalisation constant, which may be determined by
numerical integration, e.g.\ using the trapezoidal rule  
(Press et al.\ 1992). Posterior distributions for the individual
parameters are obtained by marginalization, namely integrating
over the unwanted parameters. For example, the marginal posterior
distribution for the power-law index is given by
\begin{equation}
P(\gamma |D) =\int d\sigma \, P(\sigma,\gamma |D).
\end{equation}
Parameter estimates are then obtained by taking moments of the
marginal posterior distributions.

\section{Model comparison}

Bayesian model comparison involves taking ratios of the 
posterior probabilities for models. The likelihood terms for the
models are expanded as integrals over all possible choices of
the model parameters, thereby taking account of the possibility
of different numbers of parameters in different models (e.g.\ 
Sivia \& Skilling 2006). 

More specifically, for the two models considered here -- 
the power-law plus rollover model, and the simple 
power law model -- the ratio of the posteriors
given in each case by Bayes's theorem defines the odds ratio
\begin{eqnarray}\label{eq:odds_ratio}
r_{\rm plr/pl}(D)&=&
\frac{P({\cal M}_{\rm plr}|D )}{P({\cal M}_{\rm pl}|D )} \nonumber \\
&=&\frac{P(D|{\cal M}_{\rm plr})}{P(D|{\cal M}_{\rm pl})}
\times \frac{P({\cal M}_{\rm plr})}{P({\cal M}_{\rm pl})},
\end{eqnarray}
where the models are denoted ${\cal M}_{\rm pl}$ and 
${\cal M}_{\rm plr}$ respectively. The unknown term $P(D)$ appearing
in the individual statements of Bayes's theorem for the two models
has cancelled. The two factors on the right hand side of 
equation~(\ref{eq:odds_ratio}) are given names. The ratio
\begin{equation}\label{eq:}
B_{\rm plr/pl} =  \frac{P(D|{\cal M}_{\rm plr})}
  {P(D|{\cal M}_{\rm pl })}
\end{equation}
is called the Bayes factor, and consists of the ratio of the 
likelihoods of the two models. The second ratio:
\begin{equation}
\rho_{\rm plr/pl} = \frac{P({\cal M}_{\rm plr})}{P({\cal M}_{\rm pl})},
\end{equation} 
is called the prior odds ratio. The two factors express what the 
data says about the relative probabilities of the two models, 
and what was thought before looking at the data, respectively.

For the simple power-law model, the likelihood may be expanded as an
integral over possible values of the power-law index:
\begin{equation}\label{eq:global_L_pl_gen}
P(D|{\cal M}_{\rm pl })= \int d\gamma P(D|\gamma ) P(\gamma ).
\end{equation}
A model likelihood is often referred to as a global likelihood,
when expressed in this way. Using equations~(\ref{eq:pl_likelihood}) 
and~(\ref{eq:pl_prior}) equation~(\ref{eq:global_L_pl_gen}) evaluates 
to
\begin{equation}\label{eq:global_L_pl}
P(D|{\cal M}_{\rm pl })
  =\frac{ P \left[ M+1,(\gamma_2-1)\ln \pi\right]
  -P \left[ M+1,(\gamma_1-1)\ln \pi \right]}
  {(\gamma_2-\gamma_1)S_1^{M}\pi(\ln \pi)^{M}}M!(ds)^M,
\end{equation} 
which may be evaluated numerically for given data.

Similarly for the power-law plus rollover model, which we denote
${\cal M}_{\rm plr}$, the global likelihood may be expressed as
an integral over all values of $\sigma$ and $\gamma$:
\begin{equation}\label{eq:global_L_plr_gen}
P(D|{\cal M}_{\rm plr })= \int\!\!\int d\sigma\,d\gamma\, 
  P(D|\sigma,\gamma ) P(\sigma )P(\gamma ) .
\end{equation}
Using equations~(\ref{eq:Ppost_uroll}), (\ref{eq:pl_prior}), 
and~(\ref{eq:sigma_prior}), this may be written
\begin{equation}\label{eq:global_L_plr}
P(D|{\cal M}_{\rm plr }) =\frac{(ds)^M}{\gamma_2-\gamma_1}
\int_{\sigma_1}^{\sigma_2}d\sigma 
  {\rm e}^{-S_1\Sigma/\sigma }
  \int_{\gamma_1}^{\gamma_2}d\gamma
  \frac{\sigma ^{M(\gamma-1)-1}S_1^{-M\gamma}\pi^{-\gamma}}
  {\left[\Gamma(1-\gamma)
  \left\{1-P(1-\gamma,S_1/\sigma )\right\}\right]^M},
\end{equation}
where $\sigma_1$ and $\sigma_2$ are chosen limits on the
integration in $\sigma$. The integrals may be evaluated 
numerically, e.g.\ using the trapezoidal rule.

The odds ratio for the two models may be evaluated for given data
by applying equations~(\ref{eq:global_L_pl}) 
and~(\ref{eq:global_L_plr}) in equation~(\ref{eq:odds_ratio}). 
To evaluate the odds ratio it is also necessary to make an 
assumption about the prior odds ratio $\rho_{\rm plr/pl}$. 
In many contexts it may be 
appropriate to assume the prior odds ratio is unity, and to let
the `data speak for itself.' However, this is not always the case 
[see e.g.\ discussions in Jaynes (2003) and Sivia 
and Skilling (2006)]. The resulting odds ratio may be expressed 
as evidence $\epsilon $ in decibels or dB (e.g.\ Jaynes 2003), 
according to
\begin{equation}
\epsilon = 10\log_{10}r_{\rm plr/pl}(D).
\end{equation}
Jeffreys (1961) provided a grading of the decisiveness of 
different values of the evidence. He considered values 
$|\epsilon | = 1 - 5$ as
``not worth more than a bare mention,'' 
$|\epsilon | = 5 -10$ as ``substantial,'' $|\epsilon |=10-15$ 
as ``strong,'' $|\epsilon |=15-20$ as ``very strong,'' and 
$|\epsilon |> 20$ as ``decisive.'' However, the application of 
these grades requires a decision about the choice of the prior 
odds ratio.

\clearpage

\clearpage

\begin{figure}
\plotone{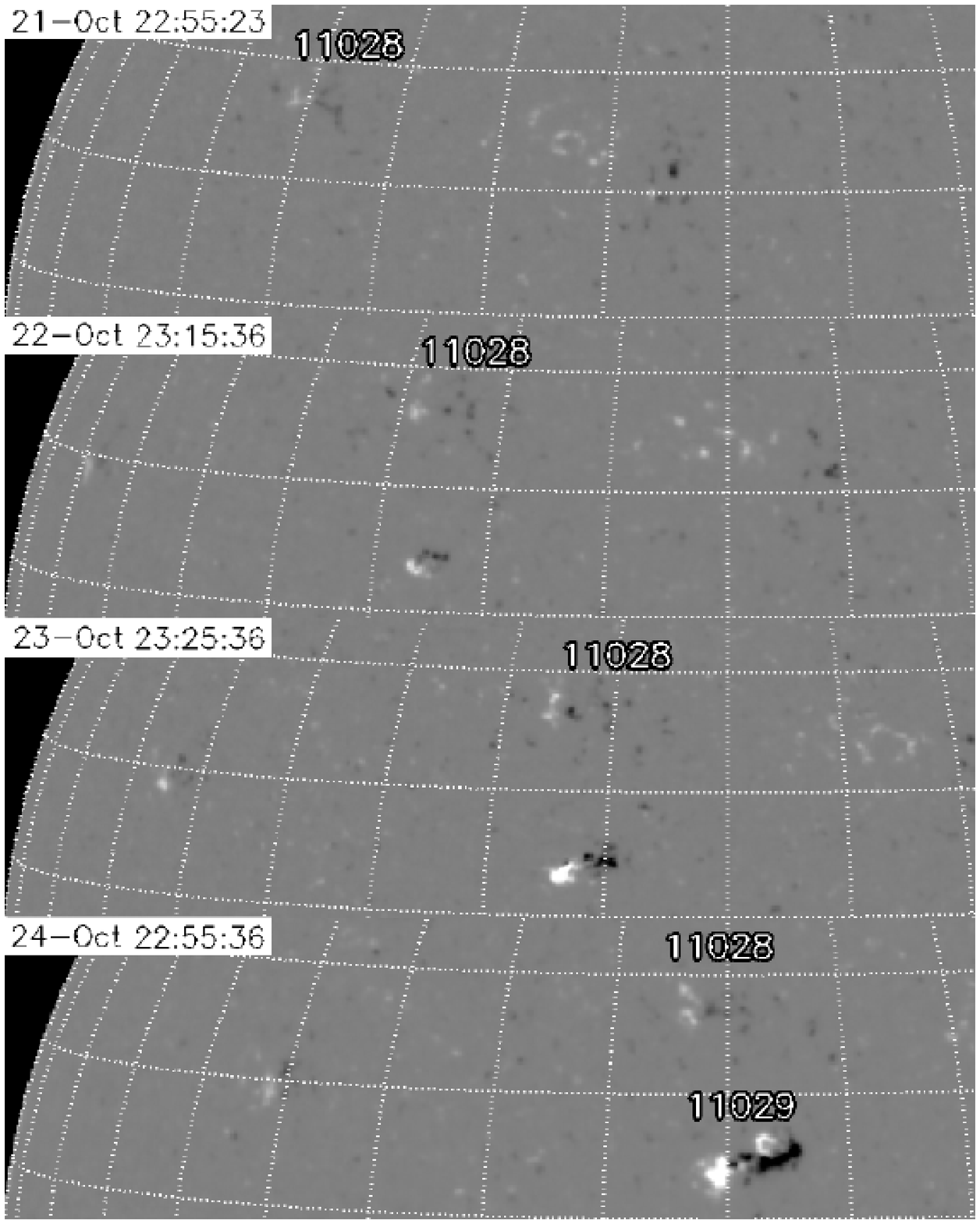}
\caption{Collage of four Global Oscillations Network Group (GONG)
magnetograms showing the emergence of AR 11029 and
its development on the disk from 21 Oct 2009 (upper panel) to
24 Oct 2009 (lower panel). Active region 11028, also shown, was
a simple $\beta$-region and did not flare.
(Images from www.solarmonitor.org.)\label{fig1}}
\end{figure}

\begin{figure}
\plotone{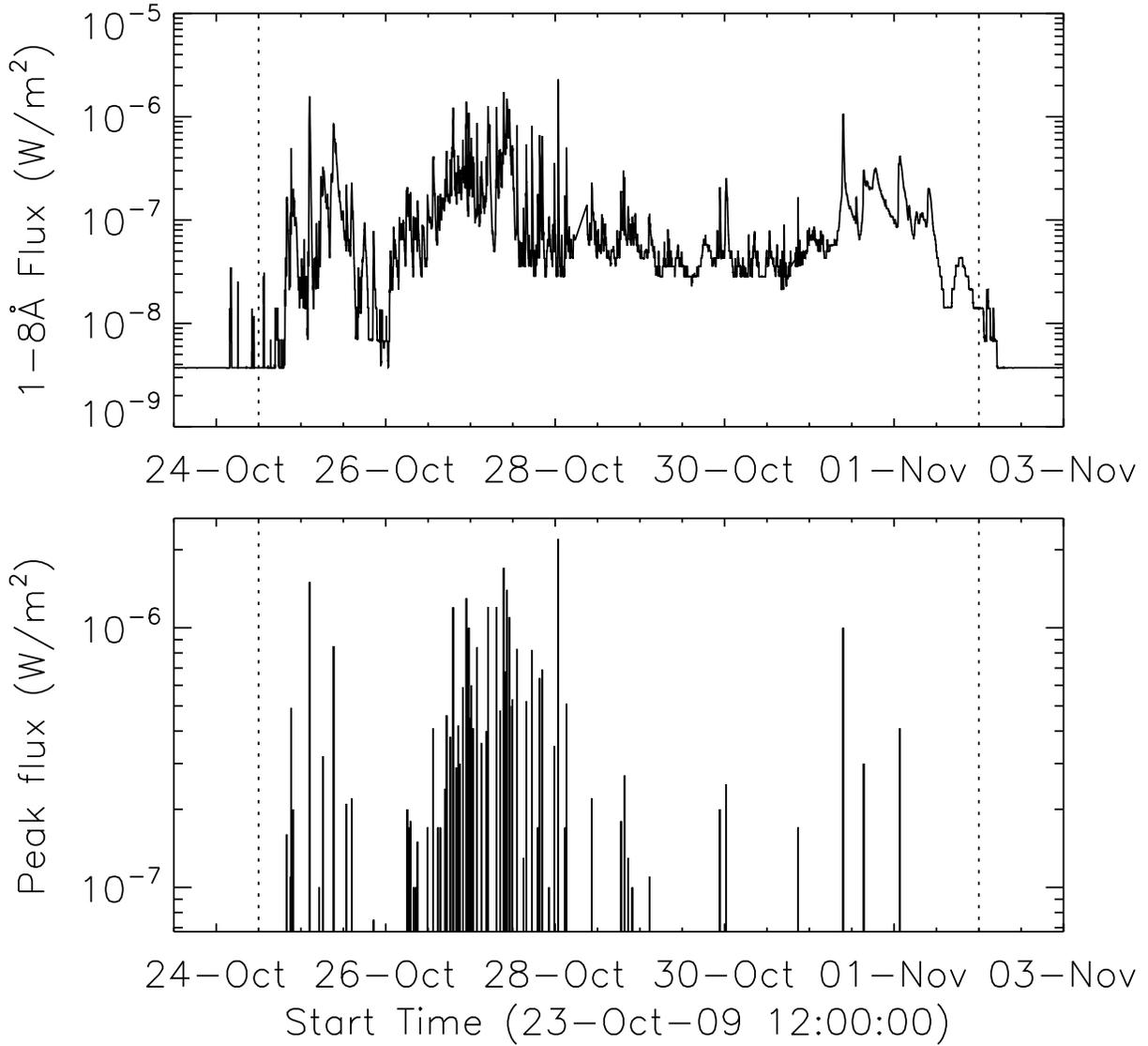}
\caption{Soft X-ray data and events for the observation interval.
The upper panel shows the one-minute Geostationary 
Observational Environmental (GOES) satellite $1$-$8\,\mbox{\AA}$
fluxes, with the 
vertical dotted lines indicating the nominal start- and end-times for
the analyses performed here. The lower panel shows the GOES events
compiled by the US Space Weather Prediction Service (SWPC), selected
from the data shown in the upper panel.\label{fig2}}
\end{figure}

\begin{figure}
\plotone{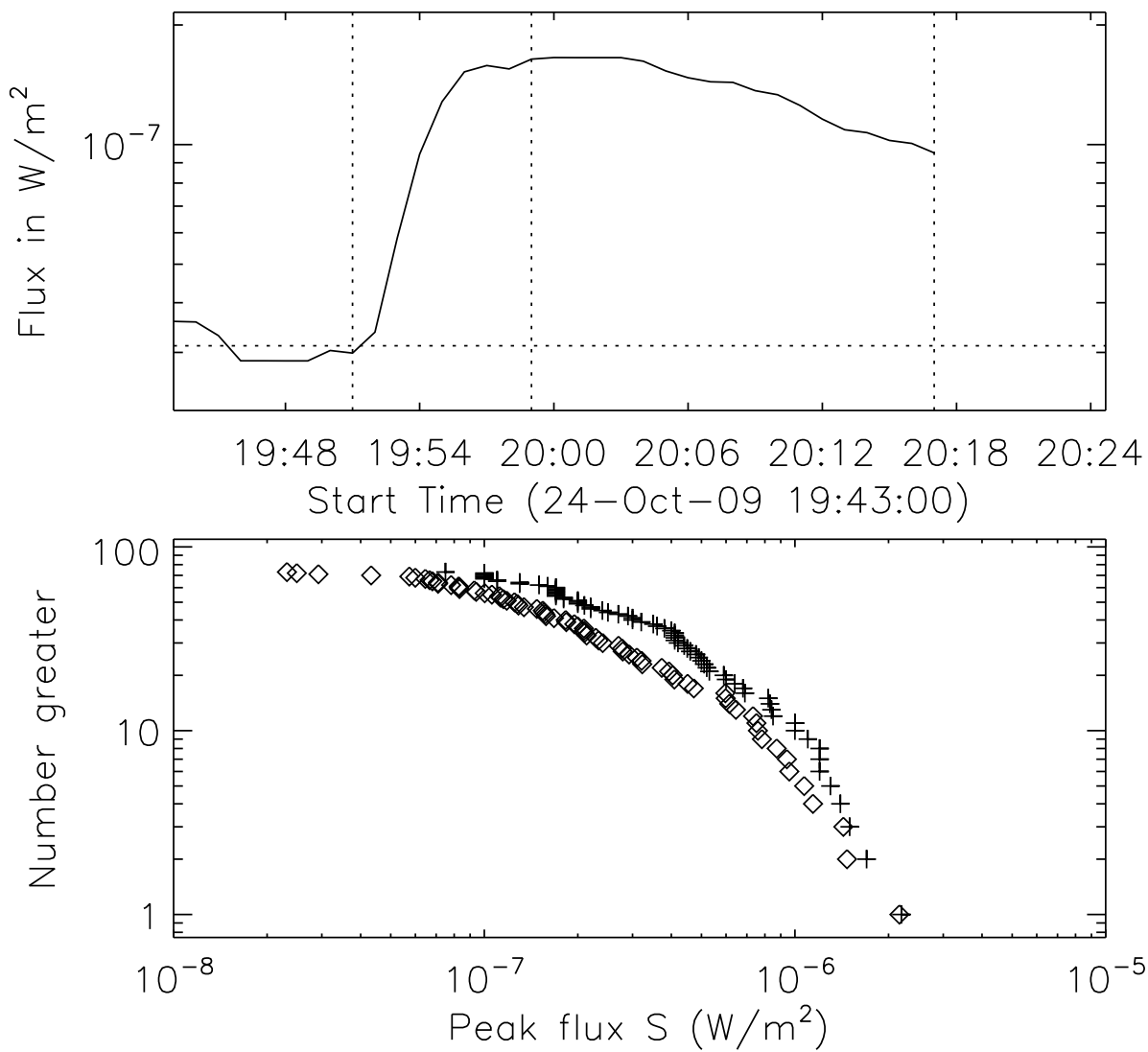}
\caption{Background subtraction.
The upper panel shows the background subtraction procedure for the 
first flare from AR 11029. The dotted vertical lines are
the start, peak, and end times for the event, and the dotted 
horizontal line is the background estimate, obtained by averaging 
the one-minute GOES $1$-$8\,\mbox{\AA}$ flux values (solid curve) for 
an interval prior to the start time equal to the rise time of
the event (peak time minus start time). The lower panel shows the 
effect of background subtraction on the cumulative number distribution 
of all events. The crosses are the peak fluxes before background 
subtraction and the diamonds are the peak fluxes after.\label{fig3}}
\end{figure}

\begin{figure}
\begin{center}
\plotone{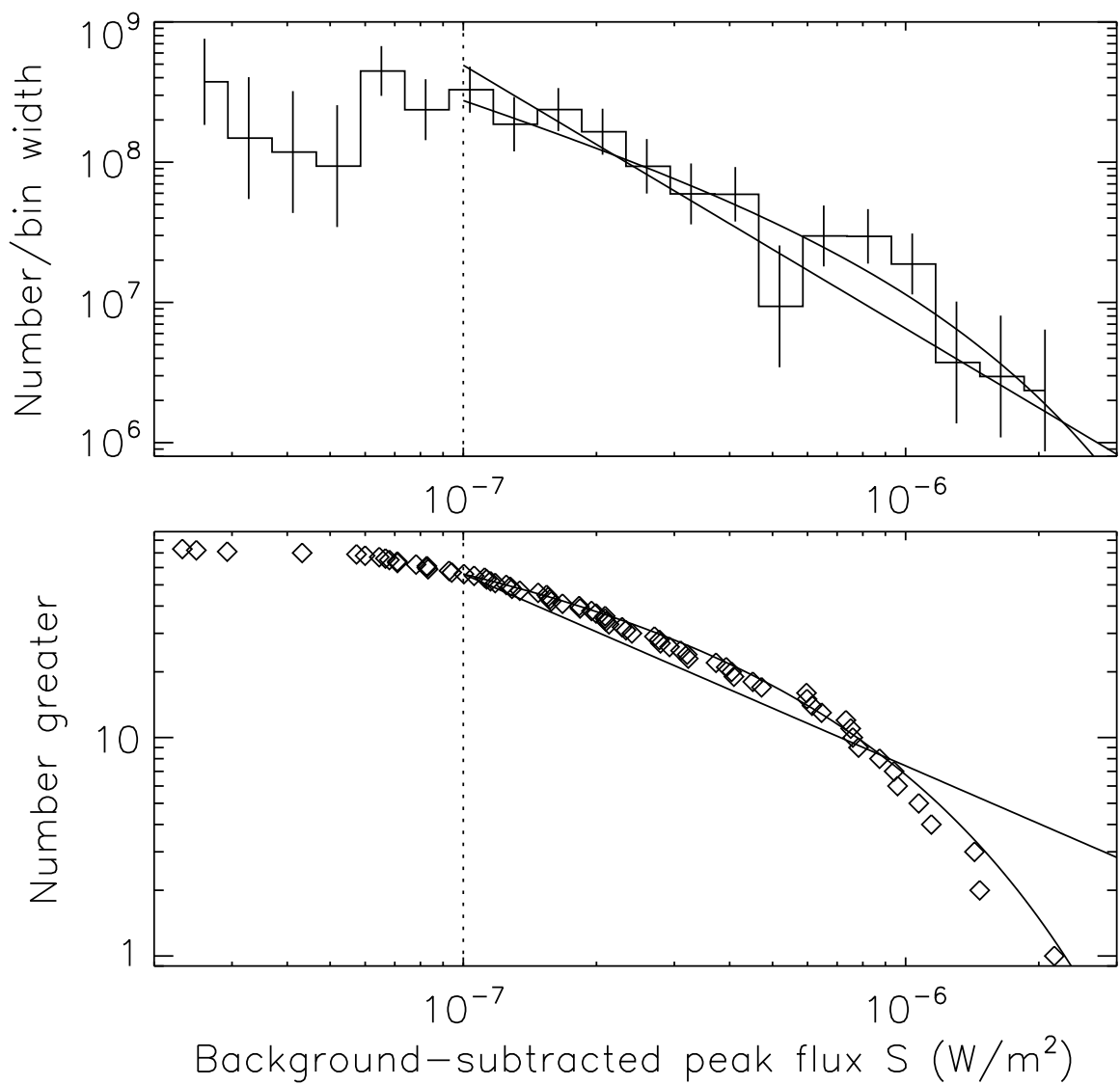}
\end{center}
\caption{Analysis of the background-subtracted peak-flux for events 
in AR 11029. The upper panel shows a histogram of the number 
distribution for peak flux, and the lower panel shows the 
cumulative number distribution (the same distribution shown in the 
lower panel of Fig.~3). The solid straight line is the inferred power-law 
model, and the solid curve is the inferred power-law plus rollover
model. The vertical dotted line shows 
$S_1=10^{-7}\,{\rm W}\,{\rm m}^{-2}$, the
flux value above which the models are assumed to apply.
The error bars in the upper panel correspond to the square root 
of the number in each bin, and are indicative only (they are not 
used in the analysis). The power law plus rollover model is found to
be much more probable: the odds ratio for the 
two models is $\approx 220$ or $23\,{\rm dB}$.
\label{fig4}}
\end{figure}

\begin{figure}
\begin{center}
\plotone{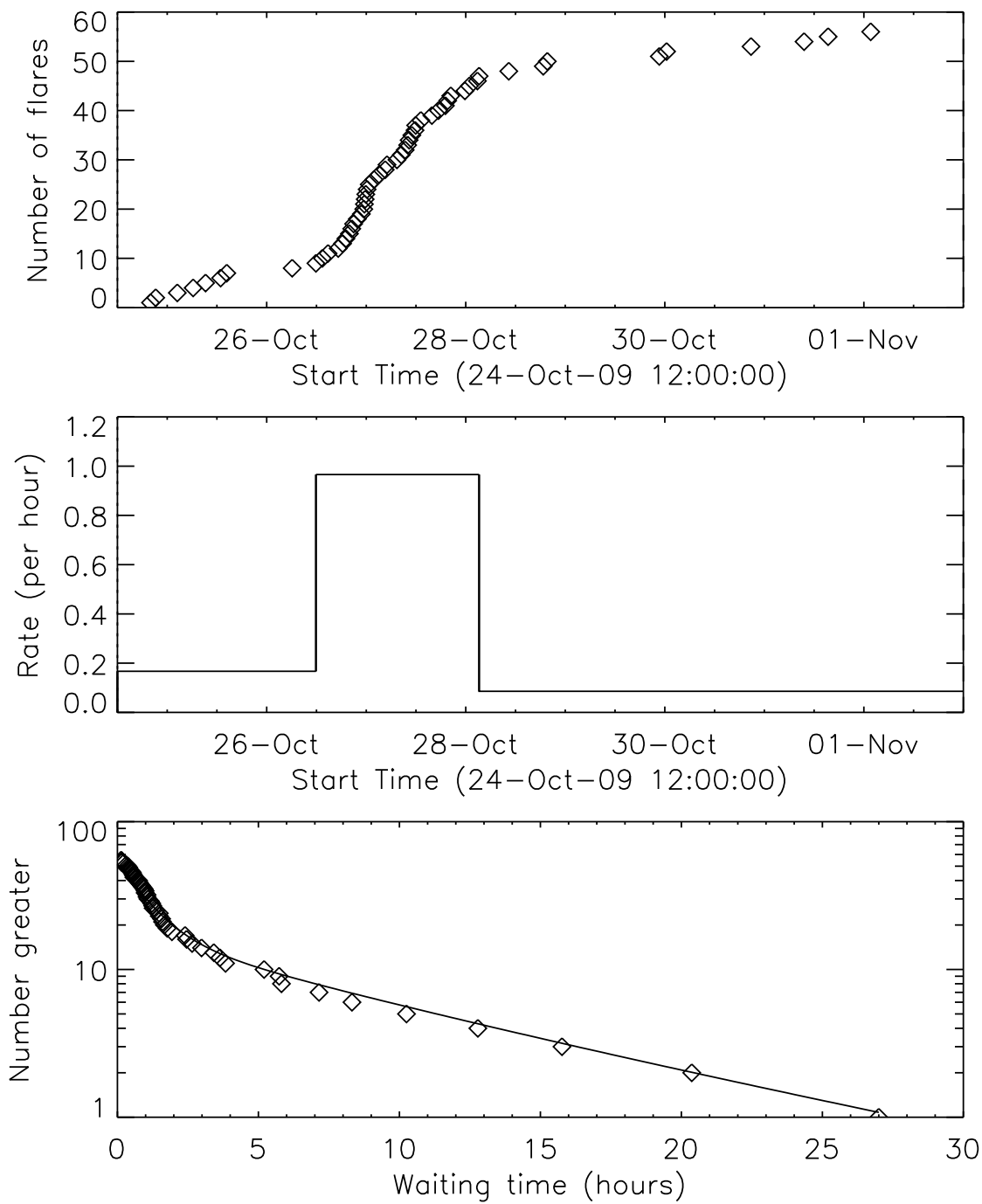}
\end{center}
\caption{Analysis of the rate of flaring in AR 11029 for the 
background-subtracted events. The upper panel 
shows the cumulative number of events versus time -- the slope 
suggests the rate. The middle panel shows the results of the 
Bayesian blocks procedure, which determines three periods with
approximately constant mean rate (three blocks). The lower panel 
shows the observed waiting-time distribution (diamonds) together 
with the piecewise-constant Poisson model corresponding to the 
Bayesian blocks analysis.\label{fig5}}
\end{figure}

\clearpage

\begin{table}
\begin{center}
\caption{Daily behavior of solar active region AR 11029.\label{tb:ar11029}}
\begin{tabular}{ccccc}
\tableline\tableline
Day & Classification\tablenotemark{a} & Sunspot area ($\mu$-hs)\tablenotemark{b} 
  & GOES events & Comments \\
\tableline
21--22 Oct & -- & -- & 0 & Emergence \\
24 Oct & $\beta$ & 50 & 4    & Sunspot formation\\
25 Oct & $\beta$ & 120 & 7  &\\
26 Oct & $\beta$--$\gamma$ & 130 & 24   &\\
27 Oct & $\beta$--$\gamma$ & 190 & 23   &\\
28 Oct & $\beta$--$\gamma$ & 260 & 8   &\\
29 Oct & $\beta$--$\gamma$ & 340 & 2   &\\
30 Oct & $\beta$ & 380 & 2   &\\
31 Oct & $\beta$ & 320 & 2   &\\
1 Nov & -- & -- & 1 & Rotated off disk  \\
\tableline
\end{tabular}
\tablenotetext{a}{Mount Wilson sunspot magnetic classifications,
obtained from Solar Region Summaries prepared by the
US National
Weather Service/National Oceanic and Atmospheric Administration
(NWS/NOAA). A $\beta$ spot is a simple bipolar spot region. A 
$\beta$--$\gamma$ spot is bipolar but more complex, such that no
single line can be drawn between two spots of opposite magnetic
polarity. (Summaries available from http://www.swpc.noaa.gov/).}
\tablenotetext{b}{Areas (in solar micro-hemispheres, $\mu$-hs)
 from the US NWS/NOAA Solar Region Summaries.}
\end{center}
\end{table}

\end{document}